\documentclass{article}
\usepackage{graphics}
\usepackage{rotating}
\input epsf.sty
\begin{document}
\newcommand{\Abstract}[2]{{\footnotesize\begin{center}ABSTRACT\end{center}
\vspace{1mm}\par#1\par
\noindent
{~}{\it #2}}}

\newcommand{\TabCap}[2]{\begin{center}\parbox[t]{#1}{\begin{center}
 \small {\spaceskip 2pt plus 1pt minus 1pt T a b l e}
 \refstepcounter{table}\thetable \\[2mm]
 \footnotesize #2 \end{center}}\end{center}}

\newcommand{\TableSep}[2]{\begin{table}[p]\vspace{#1}
\TabCap{#2}\end{table}}

\newcommand{\FigCap}[1]{\footnotesize\par\noindent Fig.\ %
 \refstepcounter{figure}\thefigure. #1\par}

\newcommand{\TableFont}{\footnotesize}
\newcommand{\TableFontIt}{\ttit}
\newcommand{\SetTableFont}[1]{\renewcommand{\TableFont}{#1}}

\newcommand{\MakeTable}[4]{\begin{table}[htb]\TabCap{#2}{#3}
 \begin{center} \TableFont \begin{tabular}{#1} #4
 \end{tabular}\end{center}\end{table}}

\newcommand{\MakeTableSep}[4]{\begin{table}[p]\TabCap{#2}{#3}
 \begin{center} \TableFont \begin{tabular}{#1} #4
 \end{tabular}\end{center}\end{table}}

\newenvironment{references}%
{
\footnotesize \frenchspacing
\renewcommand{\thesection}{}
\renewcommand{\in}{{\rm in }}
\renewcommand{\AA}{Astron.\ Astrophys.}
\newcommand{\AAS}{Astron.~Astrophys.~Suppl.~Ser.}
\newcommand{\ApJ}{Astrophys.\ J.}
\newcommand{\ApJS}{Astrophys.\ J.~Suppl.~Ser.}
\newcommand{\ApJL}{Astrophys.\ J.~Letters}
\newcommand{\AJ}{Astron.\ J.}
\newcommand{\IBVS}{IBVS}
\newcommand{\PASP}{P.A.S.P.}
\newcommand{\Acta}{Acta Astron.}
\newcommand{\MNRAS}{MNRAS}
\renewcommand{\and}{{\rm and }}
\section{{\rm REFERENCES}}
\sloppy \hyphenpenalty10000
\begin{list}{}{\leftmargin1cm\listparindent-1cm
\itemindent\listparindent\parsep0pt\itemsep0pt}}%
{\end{list}\vspace{2mm}}

\def\TYLDA{~}
\newlength{\DW}
\settowidth{\DW}{0}
\newcommand{\dw}{\hspace{\DW}}

\newcommand{\refitem}[5]{\item[]{#1} #2
\def\REFARG{#3}\ifx\REFARG\TYLDA\else, {\it#3}\fi
\def\REFARG{#4}\ifx\REFARG\TYLDA\else, {\bf#4}\fi
\def\REFARG{#5}\ifx\REFARG\TYLDA\else, {#5}\fi.}

\newcommand{\Section}[1]{\section{#1}}
\newcommand{\Subsection}[1]{\subsection{#1}}
\newcommand{\Acknow}[1]{\par\vspace{5mm}{\bf Acknowledgements.} #1}
\newcommand{\Y}{Y_\ell^m}
\newcommand{\vxi}{\mbox{\boldmath{$\xi$}}}
\newcommand{\vx}{\mbox{\boldmath{$x$}}}
\newcommand{\vnab}{\mbox{\boldmath{$\nabla$}}}

\pagestyle{myheadings}

\def\thefootnote{\fnsymbol{footnote}}
\def\th{\thinspace}

\begin{center}
{\Large\bf Mean Angular Diameters and Angular Diameter Amplitudes of Bright Cepheids\\}

\vskip3pt

{\bf P.~Moskalik$^{1}$~~ and~~ N.~A.~Gorynya$^{2}$}

\vskip3mm

{$^1$ Copernicus Astronomical Centre, ul.~Bartycka~18, 00-716 Warsaw, Poland
\\e-mail: pam@camk.edu.pl\\
$^2$ Institute of Astronomy, Russian Academy of Sciences, 48 Pyatnitskaya St.,
109017 Moscow, Russia \\e-mail: gorynya@sai.msu.ru\\}

\vskip5mm

\end{center}

\Abstract{We predict mean angular diameters and amplitudes of
angular diameter variations for all monoperiodic Population~I
Cepheids brighter than $\langle V \rangle = 8.0$~mag. The catalog
is intended to aid selecting most promising Cepheid targets for
future interferometric observations.}

\Section{Introduction}

Because of high intrinsic brightness and existence of a tight
period--luminosity relation, classical Cepheids play a central
role in building cosmological distance ladder. An accurate
calibration of their $P-L$ relation is, therefore, of fundamental
importance. While the slope of the $P-L$ relation is well
determined by Cepheids of the Large Magellanic Cloud ({\it e.g.}
Udalski et~al. 1999), the zero point is less certain. It is
usually calibrated by nearby Cepheids, whose individual distances
have to be accurately measured. This is not an easy task. Despite
several different methods being applied ({\it cf.} Fouqu\'e et~al.
2003; Feast 2003), the zero point of the $P-L$ relation remains
uncertain at $\Delta M_V = \pm 0.1$~mag level.

The advent of long-baseline interferometry offered a novel way of
Cepheid distance determination, by using purely geometrical
version of the Baade-Wesse\-link method (Lane et~al. 2000). This
approach yields the distance by comparing {\it angular} diameter
changes measured by interferometry with {\it linear} diameter
chanages inferred from observed radial velocities. So far, the
technique was succesfully applied to only five Cepheids (Lane
et~al. 2002; Kervella et~al. 2004a), but with increased resolution
of next generation of interferometers (CHARA and AMBER) more
Cepheids will become accessible.

The goal of this paper is to identify Cepheids, which are most
promising targets for observations with existing and future
interferometers. For that purpose, we calculated expected mean
angular diameters and angular diameter amplitudes for 79 brightest
monoperiodic Cepheids. Assumptions used in the calculations and
accuracy of the method are discussed in Section~2. In Section~3 we
describe the Cepheid sample and the sources of data. The results
are presented in Section~4 and conclusions of the paper are
summarized in Section~5.

\Section{Method}

The goal of this paper is to estimate for Pop.~I Cepheids the mean
angular diameters and the amplitudes of angular diameter
variations. To achieve this, we need to find for each Cepheid its
distance, mean linear radius and radius changes.

Cepheid distances were calculate with the help of the
period--luminosity relation. We adopted $P-L$ relation of Fouqu\'e
et~al. (2003):

\begin{equation}
M_V = -2.735\log P - 1.352,
\end{equation}

\noindent whose slope was determined from the LMC Cepheid sample
and the zero point was calibrated with Galactic Cepheids analysed
with infrared surface brightness method. Eq.(1) is valid for the
fundamental mode pulsators. For the first overtone Cepheids, the
observed period, $P_1$, was converted to the fundamental mode
period, $P_0$, with empirical formula

\begin{equation}
P_1/P_0 = -0.027\log P_1 + 0.716,
\end{equation}

\noindent which was derived by least square fit to periods of
Galactic double-mode Cepheids (Alcock et~al. 1995; modyfied by
Feast \& Catchpole 1997). Comparison of the absolute magnitudes,
$M_V$, given by Eq.(1) and dereddened intensity mean observed
magnitudes, $\langle V_0\rangle$, yielded the Cepheid distances.
Standard extinction coefficients were used: $A_V = 3.30\, E(B-V)$.

The mean Cepheid radii were calculated with the period--radius
relation of Gieren et~al. (1998):

\begin{equation}
\log \langle R/R_{\odot}\rangle = 0.750\log P + 1.075.
\end{equation}

\noindent This formula is essentially identical to those derived
by Laney \& Stobie (1995) and Turner \& Burke (2002). Again,
periods of first overtone Cepheids were fundamentalized with
Eq.(2).

Variation of Cepheid radius during pulsation cycle was
calculated by integrating the observed radial velocity curve, $V_r(t)$:

\begin{equation}
\Delta R(t) = -p \int_{t_0}^t [V_r(t)-\gamma]
\end{equation}

\noindent where  $\gamma = \langle V_r \rangle$ is the mean radial
velocity of the Cepheid and $p$ is the projection factor
converting observed radial velocity to pulsational velocity. For
all the Cepheids we used the same constant projection factor of
$p=1.36$ ({\it e.g.} Laney \& Stobie 1995; Kervella et~al. 2004a).

With the mean radius and the distance to the star known, the mean
angular diameter can be calculated with the formulae

\begin{equation}
\langle\theta\rangle = 9.305\ {\langle R\rangle \over d}
\end{equation}

\noindent where $\theta$ is expressed in miliarcseconds ([mas]), $R$
in units of solar radius and $d$ in parsecs. Similarly, the total
range of angular diameter variations is given by

\begin{equation}
\Delta\theta = \theta_{\rm max} -\theta_{\rm min} = 9.305\ {R_{\rm max} - R_{\rm min} \over d}.
\end{equation}

\Subsection{Accuracy of Angular Diameter Estimate}

The $P-L$ and $P-R$ relations represent Cepheids only in the
average sense. In addition to the observational scatter, both
relations have also {\it intrinsic} dispersion, which reflects
non-zero width of the Cepheid instability strip. This directly
affects accuracy of predicted Cepheid angular diameters.

When built with the reddening-independent Wesenheit index, the
$P-L$ relation for Galactic Cepheids displays scatter of 0.17\th
mag (Gieren et~al. 1998). There is a small contribution from
distance errors of individual calibrating Cepheids, which
according to Gieren et~al. are accurate to $\pm 3$\%. Taking this
into account, we find intrinsic dispersion of the $P-L$ relation
to be 0.157\th mag. This corresponds to 7.2\% uncertainty of
distances derived with Eq.~(1) and of angular diameter amplitudes
derived with Eq.~(6).

Estimating uncertainty of $\langle R\rangle$ and
$\langle\theta\rangle$ is somewhat more elaborate. We will do it
with the help of simple theoretical relations. First, we recall
that Cepheids obey well known period--mass--radius relation ({\it
e.g.} Moskalik \& Buchler 1993):

\begin{equation}
P \approx M^{-0.68}\langle R\rangle^{1.70}.
\end{equation}

\noindent Vast majority of Pop.~I Cepheids undergo core helium
burning. Stars in this evolutionary phase obey a mass--luminosity
relation:

\begin{equation}
\log\langle L\rangle = 3.55\log M + {\rm const}.
\end{equation}

\noindent We adopted here the slope derived for metalicity of
$Z=0.02$ by Alibert et~al. (1999), but evolutionary calculations
of other authors yield very similar values. Combining Eqs.~(7) and
(8) we find that {\it at a constant period}, luminosity and radius
of a Cepheid are related by

\begin{equation}
\log\langle L\rangle = 8.88\log\langle R\rangle + {\rm const}.
\end{equation}

\noindent Knowing intrinsic dispersion of the $P-L$ relation, we
find intrinsic dispersion of the $P-R$ relation to be
$\sigma(\log\langle R\rangle) = 0.007$, or equivalently
$\sigma(\langle R\rangle)/\langle R\rangle = 1.6$\%.

The estimate of mean angular diameter of a Cepheid is based on its
mean radius and its distance. It is evident, that uncertainty of
$\langle\theta\rangle$ is dominated by intrinsic dispersion of the
distance determination. However, when derived from $P-R$ and $P-L$
relations, the radius and the distance are not independent and
their inaccuracies compensate each other. Indeed, Eq.~(9) shows
that if Cepheid's radius is larger than average for its period,
its luminosity (and consequently derived distance) is also larger.
Taking this into account, we find
$\sigma(\log\langle\theta\rangle) = 0.39\,\sigma(\log\langle
L\rangle)$. Thus, 0.157\th mag intrinsic dispersion of the $P-L$
relation implies 5.6\% uncertainty of mean angular diameters
estimated with Eq.~(5).

Apart from the intrinsic width of the instability strip, the
accuracy of $\langle\theta\rangle$ and $\Delta\theta$ estimation
is also affected by observational errors. Of these, by far the
most important is the error of the colour excess $E(B-V)$, which
is $\sim 0.03$\th mag (Fernie 1990). This corresponds to the
distance uncertainity of 4.6\%. Taking into account both intrinsic
and random scatter, we find that our method should yield
$\langle\theta\rangle$ and $\Delta\theta$ accurate to 7.2\% and to
8.5\%, respectively ($1\sigma$ errors).

The error budget presented here does not account for systematical
errors, which may result from poor knowledge of $P-L$ and $P-R$
relations or of the projection factor $p$. The question of
possible systematical bias of our method will be addressed in
Section~5.1.

\Section{The Data}

For the purpose of this paper, we limited the analysis to
brightest Pop.~I Cepheids, specifically to those with $\langle V
\rangle < 8.0$\th mag. The starting source was the online DDO
Database of Galactic Classical Cepheids (Fernie et~al. 1995),
which includes 81 stars satisfying our brightness criterion. We
supplemented this catalog with three recently discovered bright
Cepheids: CK~Cam, V898~Cen and V411~Lac. We excluded from the list
four double-mode variables (CO~Aur, TU~Cas, EW~Sct, and U~TrA) as
well as a peculiar variable amplitude Cepheid V473~Lyr. Because of
complicated form of pulsations, these stars are not suitable
targets for interferometric investigation. Our final sample
contains 79 objects.

In our analysis, we adopted intensity mean magnitudes $\langle
V\rangle$ and colour excesses $E(B-V)$ given by DDO Database. The
latter are defined on a uniform scale of Fernie (1990). For
$\alpha$~UMi and V1334~Cyg, colour excesses in DDO Database are
negative. We assumed $E(B-V)=0$ for these two stars. For CK~Cam
and V411~Lac, the values of $\langle V\rangle$ were taken from
Berdnikov et~al. (2000) and from Groenewegen \& Oudmaijer (2000),
respectively. In case of V898~Cen, we determined $\langle
V\rangle$ directly from published photometry (Berdnikov et~al.
1999; Berdnikov \& Caldwell 2001; Berdnikov \& Turner 2001). The
colour excesses of CK~Cam and V898~Cen were calculated with
formula of Fernie (1994), which puts them on the same scale as
used in DDO Database. For V411~Lac, $E(B-V)$ was determined by
comparing observed $(B-V)$ colour (Groenewegen \& Oudmaijer 2000)
with $(B-V)_0$, predicted by period--colour relation of Laney \&
Stobie (1994).

\begin{table}
\caption{Predicted Angular Diameters of Bright Classical Cepheids}
\begin{center}
\begin{tabular}{lccccccccc}
\hline
\noalign{\smallskip}
Star          &    & $\log P$
                           & $\langle V\rangle$
                                   & $E(B-V)$
                                           & d          & $\langle R\rangle$
                                                                      & $\Delta R$   & $\langle\theta\rangle$
                                                                                             & $\Delta\theta$ \\
\noalign{\smallskip}
\hline
\noalign{\smallskip}
 $\ell$ Car   &    & 1.551 & 3.724 & 0.170 & \hfill 564 &       173.0 &       33.060 & 2.854 &     0.545 \\
 SV Vul       &    & 1.653 & 7.220 & 0.570 &       1748 &       206.5 &       50.755 & 1.099 &     0.270 \\
 U Car        &    & 1.588 & 6.288 & 0.283 &       1622 &       184.7 &       44.013 & 1.059 &     0.252 \\
 RS Pup       &    & 1.617 & 6.947 & 0.446 &       1778 &       193.9 &       47.730 & 1.015 &     0.250 \\
 $\eta$ Aql   &    & 0.856 & 3.897 & 0.149 & \hfill 263 & \hfill 52.1 & \hfill 6.386 & 1.845 &     0.226 \\
 T Mon        &    & 1.432 & 6.124 & 0.209 &       1382 &       140.9 &       32.557 & 0.949 &     0.219 \\
 $\beta$ Dor  &    & 0.993 & 3.731 & 0.044 & \hfill 339 & \hfill 66.0 & \hfill 7.823 & 1.810 &     0.214 \\
 X Cyg        &    & 1.214 & 6.391 & 0.288 &       1054 & \hfill 96.8 &       20.870 & 0.855 &     0.184 \\
 $\delta$ Cep &    & 0.730 & 3.954 & 0.092 & \hfill 251 & \hfill 41.9 & \hfill 4.870 & 1.554 &     0.181 \\
 RZ Vel       &    & 1.310 & 7.079 & 0.335 &       1519 &       114.1 &       27.793 & 0.699 &     0.170 \\
 $\zeta$ Gem  &    & 1.006 & 3.918 & 0.018 & \hfill 391 & \hfill 67.6 & \hfill 6.712 & 1.607 &     0.160 \\
 TT Aql       &    & 1.138 & 7.141 & 0.495 & \hfill 988 & \hfill 84.9 &       16.721 & 0.800 &     0.158 \\
 W Sgr        &    & 0.881 & 4.668 & 0.111 & \hfill 410 & \hfill 54.4 & \hfill 6.632 & 1.235 &     0.151 \\
 X Sgr        &    & 0.846 & 4.549 & 0.197 & \hfill 326 & \hfill 51.2 & \hfill 4.790 & 1.463 &     0.137 \\
 Y Oph        &    & 1.234 & 6.169 & 0.655 & \hfill 558 &       100.1 & \hfill 8.044 & 1.668 &     0.134 \\
 VY Car       &    & 1.279 & 7.443 & 0.243 &       1986 &       108.1 &       23.831 & 0.507 &     0.112 \\
 S Sge        &    & 0.923 & 5.622 & 0.127 & \hfill 655 & \hfill 58.5 & \hfill 7.177 & 0.832 &     0.102 \\
 Y Sgr        &    & 0.761 & 5.744 & 0.205 & \hfill 502 & \hfill 44.3 & \hfill 5.327 & 0.821 &     0.099 \\
 U Aql        &    & 0.847 & 6.446 & 0.399 & \hfill 575 & \hfill 51.3 & \hfill 5.994 & 0.830 &     0.097 \\
 U Vul        &    & 0.903 & 7.128 & 0.654 & \hfill 573 & \hfill 56.5 & \hfill 5.574 & 0.917 &     0.091 \\
 U Sgr        &    & 0.829 & 6.695 & 0.403 & \hfill 626 & \hfill 49.7 & \hfill 5.741 & 0.739 &     0.085 \\
 S Nor        &    & 0.989 & 6.394 & 0.189 & \hfill 924 & \hfill 65.6 & \hfill 7.809 & 0.661 &     0.079 \\
 S Mus        &    & 0.985 & 6.118 & 0.147 & \hfill 863 & \hfill 65.1 & \hfill 7.147 & 0.703 &     0.077 \\
 RX Cam       &    & 0.898 & 7.682 & 0.569 & \hfill 837 & \hfill 56.1 & \hfill 6.776 & 0.623 &     0.075 \\
 AW Per       &    & 0.810 & 7.492 & 0.534 & \hfill 724 & \hfill 48.2 & \hfill 5.728 & 0.619 &     0.074 \\
 RT Aur       &    & 0.571 & 5.446 & 0.051 & \hfill 435 & \hfill 31.9 & \hfill 3.308 & 0.682 &     0.071 \\
 W Gem        &    & 0.898 & 6.950 & 0.283 & \hfill 923 & \hfill 56.1 & \hfill 7.084 & 0.566 &     0.071 \\
 R Mus        &    & 0.876 & 6.298 & 0.120 & \hfill 851 & \hfill 53.9 & \hfill 6.184 & 0.590 & $\>$0.068:\\
 AX Cir       &    & 0.722 & 5.880 & 0.153 & \hfill 550 & \hfill 41.4 & \hfill 3.933 & 0.700 &     0.067 \\
 V Cen        &    & 0.740 & 6.836 & 0.289 & \hfill 711 & \hfill 42.6 & \hfill 5.096 & 0.559 &     0.067 \\
 T Vul        &    & 0.647 & 5.754 & 0.064 & \hfill 541 & \hfill 36.3 & \hfill 3.656 & 0.625 &     0.063 \\
 YZ Sgr       &    & 0.980 & 7.358 & 0.292 &       1217 & \hfill 64.6 & \hfill 7.740 & 0.494 &     0.059 \\
 RV Sco       &    & 0.783 & 7.040 & 0.342 & \hfill 760 & \hfill 45.9 & \hfill 4.716 & 0.562 &     0.058 \\
 R Cru        &    & 0.765 & 6.766 & 0.192 & \hfill 823 & \hfill 44.6 & \hfill 4.986 & 0.504 &     0.056 \\
 S TrA        &    & 0.801 & 6.397 & 0.100 & \hfill 835 & \hfill 47.4 & \hfill 4.912 & 0.528 &     0.055 \\
 XX Cen       &    & 1.040 & 7.818 & 0.260 &       1702 & \hfill 71.6 &       10.016 & 0.391 &     0.055 \\
 S Cru        &    & 0.671 & 6.600 & 0.163 & \hfill 708 & \hfill 37.9 & \hfill 4.124 & 0.498 &     0.054 \\
 V636 Sco     &    & 0.832 & 6.654 & 0.217 & \hfill 819 & \hfill 50.0 & \hfill 4.766 & 0.568 &     0.054 \\
 RX Aur       &    & 1.065 & 7.655 & 0.276 &       1592 & \hfill 74.8 & \hfill 9.167 & 0.437 &     0.054 \\
 BB Sgr       &    & 0.822 & 6.947 & 0.284 & \hfill 836 & \hfill 49.1 & \hfill 4.779 & 0.547 &     0.053 \\
\noalign{\smallskip}
\hline
\end{tabular}
\end{center}
\end{table}

\setcounter{table}{0}
\begin{table}
\caption{ -- {\it continued}}
\begin{center}
\begin{tabular}{lccccccccc}
\hline
\noalign{\smallskip}
Star          &    & $\log P$
                           & $\langle V\rangle$
                                   & $E(B-V)$
                                           & d          & $\langle R\rangle$
                                                                      & $\Delta R$   & $\langle\theta\rangle$
                                                                                             & $\Delta\theta$ \\
\noalign{\smallskip}
\hline
\noalign{\smallskip}
 T Cru        &    & 0.828 & 6.566 & 0.193 & \hfill 811 & \hfill 49.7 & \hfill 4.279 & 0.570 &     0.049 \\
 AP Sgr       &    & 0.704 & 6.955 & 0.192 & \hfill 831 & \hfill 40.1 & \hfill 4.419 & 0.449 &     0.049 \\
 V350 Sgr     &    & 0.712 & 7.483 & 0.312 & \hfill 893 & \hfill 40.7 & \hfill 4.510 & 0.424 &     0.047 \\
 ER Car       &    & 0.888 & 6.824 & 0.101 &       1132 & \hfill 55.0 & \hfill 5.311 & 0.452 &     0.044 \\
 BG Vel       &    & 0.840 & 7.635 & 0.448 & \hfill 915 & \hfill 50.7 & \hfill 4.266 & 0.516 & $\>$0.043:\\
 FF Aql       & FO & 0.650 & 5.372 & 0.224 & \hfill 434 & \hfill 47.8 & \hfill 1.983 & 1.024 &     0.042 \\
 SU Cyg       &    & 0.585 & 6.859 & 0.096 & \hfill 792 & \hfill 32.6 & \hfill 3.556 & 0.383 &     0.042 \\
 CK Cam       &    & 0.518 & 7.544 & 0.457 & \hfill 577 & \hfill 29.1 & \hfill 2.626 & 0.469 &     0.042 \\
 BF Oph       &    & 0.609 & 7.337 & 0.247 & \hfill 809 & \hfill 34.0 & \hfill 3.438 & 0.391 &     0.040 \\
 AP Pup       &    & 0.706 & 7.371 & 0.208 & \hfill 985 & \hfill 40.2 & \hfill 4.084 & 0.380 &     0.039 \\
 V Vel        &    & 0.641 & 7.589 & 0.209 &       1002 & \hfill 35.9 & \hfill 4.168 & 0.334 &     0.039 \\
 V381 Cen     &    & 0.706 & 7.653 & 0.205 &       1127 & \hfill 40.2 & \hfill 4.454 & 0.332 & $\>$0.037:\\
 V482 Sco     &    & 0.656 & 7.965 & 0.360 & \hfill 965 & \hfill 36.9 & \hfill 3.826 & 0.356 &     0.037 \\
 V Car        &    & 0.826 & 7.362 & 0.174 &       1201 & \hfill 49.5 & \hfill 4.665 & 0.383 &     0.036 \\
 V636 Cas     &    & 0.923 & 7.199 & 0.700 & \hfill 566 & \hfill 58.5 & \hfill 2.134 & 0.961 &     0.035 \\
 AT Pup       &    & 0.824 & 7.957 & 0.183 &       1555 & \hfill 49.3 & \hfill 5.929 & 0.295 & $\>$0.035:\\
 R TrA        &    & 0.530 & 6.660 & 0.127 & \hfill 644 & \hfill 29.7 & \hfill 2.207 & 0.429 &     0.032 \\
 V1344 Aql    &    & 0.874 & 7.767 & 0.574 & \hfill 837 & \hfill 53.7 & \hfill 2.873 & 0.597 &     0.032 \\
 V496 Aql     &    & 0.833 & 7.751 & 0.413 &       1008 & \hfill 50.1 & \hfill 3.390 & 0.462 &     0.031 \\
 AH Vel       & FO & 0.626 & 5.695 & 0.074 & \hfill 614 & \hfill 45.8 & \hfill 1.931 & 0.695 &     0.029 \\
 V1162 Aql    &    & 0.730 & 7.798 & 0.205 &       1242 & \hfill 42.0 & \hfill 3.817 & 0.314 &     0.029 \\
 SZ Tau       & FO & 0.498 & 6.531 & 0.294 & \hfill 548 & \hfill 36.6 & \hfill 1.561 & 0.621 &     0.027 \\
 SU Cas       & FO & 0.290 & 5.970 & 0.287 & \hfill 328 & \hfill 25.4 & \hfill 0.916 & 0.721 &     0.026 \\
 MY Pup       & FO & 0.756 & 5.677 & 0.064 & \hfill 730 & \hfill 57.5 & \hfill 1.566 & 0.734 &     0.020 \\
 V659 Cen     & FO & 0.750 & 6.598 & 0.134 & \hfill 995 & \hfill 57.0 & \hfill 2.135 & 0.533 &     0.020 \\
 GH Lup       &    & 0.967 & 7.635 & 0.364 &       1220 & \hfill 63.2 & \hfill 2.561 & 0.482 &     0.020 \\
 DT Cyg       & FO & 0.398 & 5.774 & 0.039 & \hfill 501 & \hfill 30.7 & \hfill 0.944 & 0.570 &     0.018 \\
 BG Cru       & FO & 0.524 & 5.487 & 0.053 & \hfill 505 & \hfill 38.3 & \hfill 0.930 & 0.705 &     0.017 \\
 IR Cep       & FO & 0.325 & 7.784 & 0.434 & \hfill 632 & \hfill 27.0 & \hfill 1.122 & 0.398 &     0.017 \\
 V950 Sco     & FO & 0.529 & 7.302 & 0.267 & \hfill 847 & \hfill 38.6 & \hfill 1.472 & 0.424 &     0.016 \\
 AV Cir       & FO & 0.486 & 7.439 & 0.397 & \hfill 701 & \hfill 35.9 & \hfill 1.185 & 0.476 &     0.016 \\
 $\alpha$ UMi & FO & 0.599 & 1.982 & 0.000 & \hfill 120 & \hfill 43.7 & \hfill 0.173 & 3.388 &     0.013 \\
 V440 Per     & FO & 0.879 & 6.282 & 0.273 & \hfill 822 & \hfill 71.5 & \hfill 1.066 & 0.809 &     0.012 \\
 BP Cir       & FO & 0.380 & 7.560 & 0.235 & \hfill 828 & \hfill 29.7 & \hfill 1.047 & 0.334 &     0.012 \\
 V1334 Cyg    & FO & 0.523 & 5.871 & 0.000 & \hfill 653 & \hfill 38.2 & \hfill 0.797 & 0.545 &     0.011 \\
 V411 Lac     & FO & 0.464 & 7.860 & 0.171 &       1166 & \hfill 34.5 & \hfill 1.326 & 0.275 &     0.011 \\
 V737 Cen     &    & 0.849 & 6.719 & 0.216 & \hfill 863 & \hfill 51.5 &         ---  & 0.555 &      ---  \\
 LR TrA       & FO & 0.385 & 7.808 & 0.281 & \hfill 871 & \hfill 30.0 &         ---  & 0.321 &      ---  \\
 V898 Cen     & FO & 0.547 & 7.959 & 0.000 &       1762 & \hfill 39.9 &         ---  & 0.211 &      ---  \\
\noalign{\smallskip}
\hline
\end{tabular}
\end{center}
\end{table}

The radial velocity data were collected from literature and
supplemented, when needed, by unpublished data available to the
authors. No $V_r$ measurements were found for V737~Cen, V898~Cen
and LR~TrA. Several Cepheids display orbital velocity variations.
This is the case for U~Aql, FF~Aql, V496~Aql, RX~Cam, XX~Cen,
AX~Cir, BP~Cir, SU~Cyg, V1334~Cyg, T~Mon, S~Mus, AW~Per, S~Sge,
W~Sgr, V350~Sgr, V636~Sco, U~Vul and $\alpha$~UMi. For these
stars, orbital motion was removed before pulsation velocity curve
was built. We refer the reader to Moskalik et~al. (2005) for
detailed discussion of this procedure, as well as for the list of
velocity data used in the current paper. We would like to stress
that the list of binaries given above is not intended to be
complete. Several other Cepheids are likely binaries, {\it e.g.}
U~Car and T~Cru (Bersier 2002) or X~Sgr (Szabados 1990), but their
orbital motion does not show up in the data used here.

Our Cepheid sample contains eighteen overtone pulsators. Except of
$\alpha$~UMi, they have all been identified by Fourier
decomposition of their lightcurves (Antonello et~al. 1990;
Zakrzewski et~al. 2000) or radial velocity curves (Kienzle et~al.
1999, 2000; Moskalik et~al. 2005). The overtone nature of Polaris
was first established by Feast \& Catchpole (1997) on the basis of
Hipparcos paralax. It was later confirmed with different methods
by Moskalik \& Og{\l}oza (2000) and by Nordgren et~al. (2000).

\Section{Results}

Results of our calculations are summarized in Table~1. For each
Cepheid we list logarithm of observed period $\log P$ in [d],
intensity mean magnitude $\langle V\rangle$ and colour excess
$E(B-V)$, both in [mag], inferred distance $d$ in [pc], mean
radius $\langle R\rangle$ and full amplitude of radius variations
$\Delta R = R_{\rm max}-R_{\rm min}$, both in units of
$R_{\odot}$, and mean angular diameter $\langle\theta\rangle$ and
full amplitude of angular diameter variations $\Delta\theta$, both
in [mas]. First overtone pulsators are marked with symbol FO
placed next to the Cepheid's name. The stars are ordered by
decreasing $\Delta\theta$.

For three Ceheids listed at the bottom of Table~1, $\Delta R$ and
$\Delta\theta$ cannot be calculated because of lack of radial
velocity data. For these stars only rough estimates can be given.
From our Cepheid sample we find $\Delta R/\langle R\rangle =
0.020-0.045$ for the overtone pulsators and $\Delta R/\langle
R\rangle = 0.070-0.125$ for the fundamental mode pulsators with
$\log P \sim 0.85$. On this basis, we estimate angular diameter
aplitudes to be in the range of $0.039-0.069$\th mas for V737~Cen,
$0.006-0.014$\th mas for LR~TrA and $0.004-0.009$\th mas for
V898~Cen.

\Subsection{Comparison with Observations}

It is instructive to compare Cepheid angular diameters predicted
in Table~1 with those determined from actual interferometric
observations. So far, mean angular diameters were measured for
nine Pop.~I Cepheids, but angular diameter variability was
detected only in five of them. These observational results are
summarized in Table~2. The values of $\langle\theta\rangle$ (and
their errors) were taken from Kervella et~al. (2004b), except of
$\alpha$~UMi, for which result of Nordgren et~al. (2000) is
listed. The amplitudes of angular diameter variations,
$\Delta\theta$, are usually not given in the papers. We recovered
them from plots of Kervella et~al. (2004a) and Lane et~al. (2002).
For $\eta$~Aql, weighted mean of the two measurements is given. In
all cases we assumed, somewhat arbitrarily, that the error of
$\Delta\theta$ determination is the same as the corresponding
error of $\langle\theta\rangle$.

\begin{table}
\caption{Observed Angular Diameters of Cepheids}
\begin{center}
\begin{tabular}{lccc}
\hline
\noalign{\smallskip}
Star          & $\log P$
                      & $\langle\theta_{\rm LD}\rangle$
                                          & $\Delta\theta_{\rm LD}$ \\
\noalign{\smallskip}
\hline
\noalign{\smallskip}
 $\alpha$ UMi & 0.599 & 3.280 $\pm$ 0.020 & ------            \\
 $\delta$ Cep & 0.730 & 1.521 $\pm$ 0.010 & ------            \\
 X Sgr        & 0.846 & 1.471 $\pm$ 0.033 & ------            \\
 $\eta$ Aql   & 0.856 & 1.791 $\pm$ 0.022 & 0.212 $\pm$ 0.026 \\
 W Sgr        & 0.881 & 1.312 $\pm$ 0.029 & 0.163 $\pm$ 0.029 \\
 $\beta$ Dor  & 0.993 & 1.884 $\pm$ 0.024 & 0.207 $\pm$ 0.024 \\
 $\zeta$ Gem  & 1.006 & 1.688 $\pm$ 0.022 & 0.179 $\pm$ 0.030 \\
 Y Oph        & 1.234 & 1.438 $\pm$ 0.051 & ------            \\
 $\ell$ Car   & 1.551 & 2.988 $\pm$ 0.012 & 0.529 $\pm$ 0.012 \\
\noalign{\smallskip}
\hline
\end{tabular}
\end{center}
NOTE -- $\langle\theta_{\rm LD}\rangle$ and $\Delta\theta_{\rm LD}$
(in [mas]) are limb darkened angular diameters, see {\it e.g.}
Kervella et~al. (2004a).
\end{table}

\begin{figure}[p]
\centering
\vskip -3.0truecm
\resizebox*{18cm}{18cm}{\includegraphics{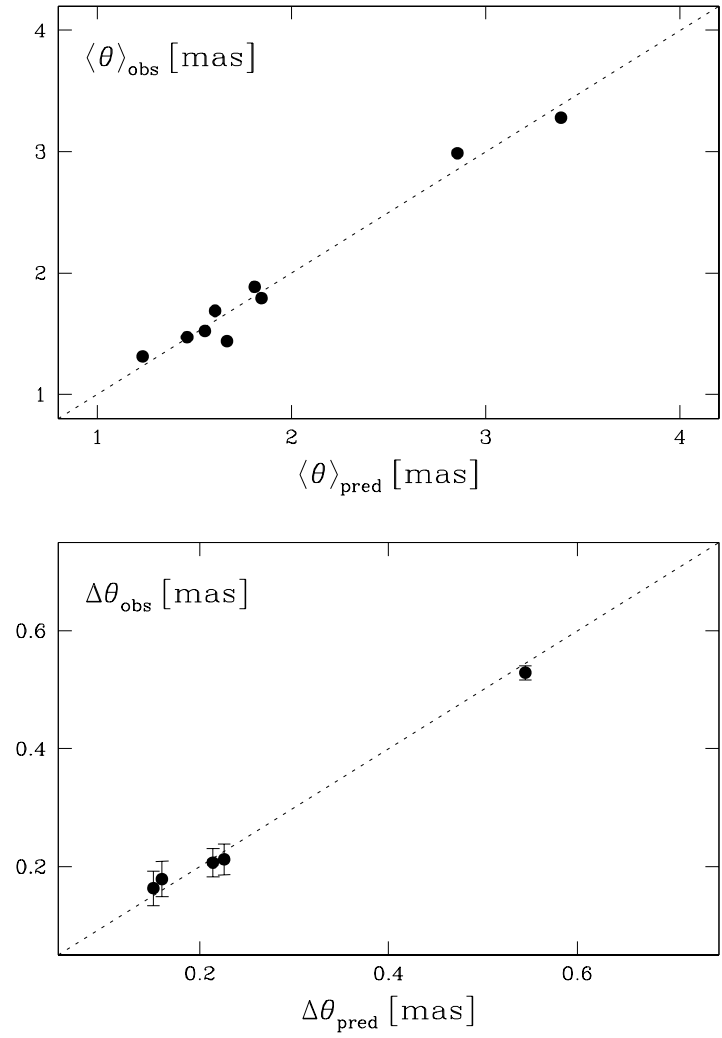}}
\vskip -4.0truecm
\caption{Observed {\it vs.} predicted mean angular diameters (top)
         and angular diameter amplitudes (bottom) for Cepheids of
         Table~2. Error bars of $\langle\theta\rangle$ are smaller
         than the symbols. The dotted lines have slope of unity and
         are not fits to the data.}
\label{f1}
\end{figure}

\begin{figure}
\vskip -3.0truecm
\resizebox*{18cm}{18cm}{\includegraphics{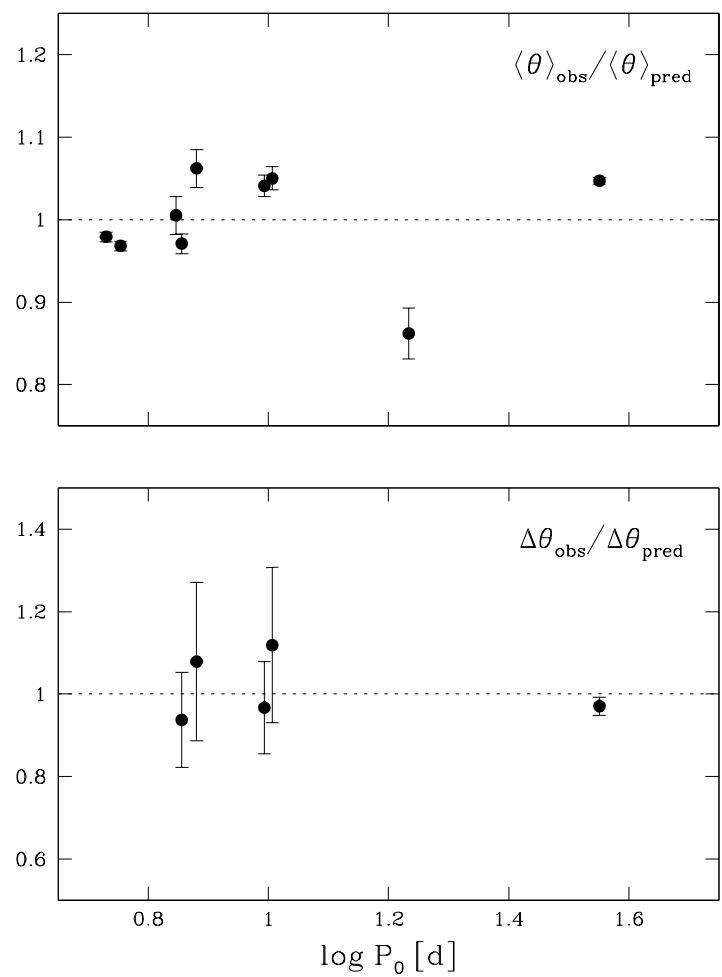}}
\vskip -4.0truecm
\caption{$\langle\theta\rangle_{\rm obs}/\langle\theta\rangle_{\rm pred}$
        (top) and $\Delta\theta_{\rm obs}/\Delta\theta_{\rm pred}$
        (bottom) {\it vs.} fundamental mode period for Cepheids of
        Table~2. Period of $\alpha$~UMi was fundamentalized with Eq.(2)}
\label{f2}
\end{figure}

In Fig.\th\ref{f1} we plot observed {\it vs.} predicted values of
$\langle\theta\rangle$ and $\Delta\theta$ for Cepheids of Table~2.
A very good overall agreement is evident. Ratios of
observed-to-predicted values of $\langle\theta\rangle$ and of
$\Delta\theta$ are ploted {\it vs.} fundamental mode period in
Fig.\th\ref{f2}. The ratios show no trends with the pulsation
period. Predicted angular diameter amplitudes, $\Delta\theta_{\rm
pred}$, differ from the observed ones by no more than $1.3\sigma$.
The weighted mean of $\Delta\theta_{\rm obs}/\Delta\theta_{\rm
pred}$ ratio is

$$\overline{\Delta\theta_{\rm obs}/\Delta\theta_{\rm pred}} = 0.973 \pm 0.021.$$

\noindent In case of $\langle\theta\rangle_{\rm
obs}/\langle\theta\rangle_{\rm pred}$, a statistically significant
scatter of $\sigma=0.042$ is seen. This is not unexpeced and
reflects intrinsic dispersion of $P-L$ and $P-R$ relations used to
estimate $\langle\theta\rangle$. The weighted mean of
$\langle\theta\rangle_{\rm obs}/\langle\theta\rangle_{\rm pred}$
ratio is

$$\overline{\langle\theta\rangle_{\rm obs}/\langle\theta\rangle_{\rm pred}} = 1.011 \pm 0.014.$$

\noindent We conclude, that the method outlined in Section~2
yields estimates of $\langle\theta\rangle$ and of $\Delta\theta$,
which are statistically unbiased and in good agreement with the
observations across the entire range of pulsation periods.


\Subsection{Prospective Targets for Interferometric Observations}

\begin{figure}
\vskip -3.0truecm
\resizebox*{18cm}{18cm}{\includegraphics{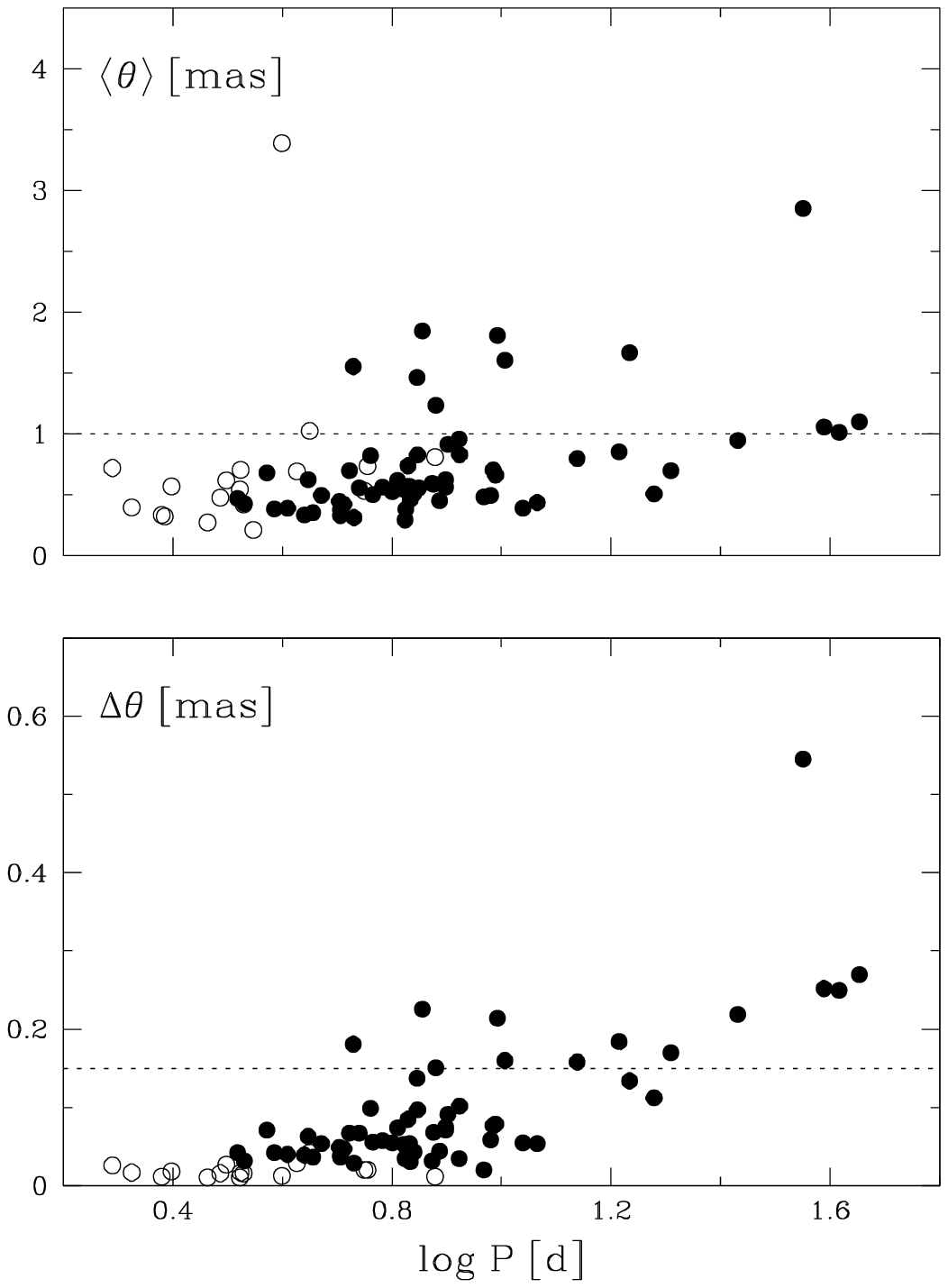}}
\vskip -4.0truecm
\caption{Predicted mean angular diameters (top) and full amplitudes of
         angular diameter variations (bottom) for Classical Cepheids
         brighter than $\langle V\rangle = 8.0$~mag. Fundamental and
         overtone pulsators are ploted with filled and open circles,
         respectively. Reference values of $\langle\theta\rangle =1.0$~mas
         and $\Delta\theta = 0.15$\th mas (see text) are marked with dotted
         lines.}
\label{f3}
\end{figure}

In Fig.\th\ref{f3} we display $\langle\theta\rangle$ and
$\Delta\theta$ {\it vs.} pulsation period for all Cepheids of our
sample. Fundamental mode and overtone pulsators are ploted as
filled and open circles, respectively.

At currently demonstrated level of technology, the achievable
accuracy of $\langle\theta\rangle$ determination is about
0.01\thinspace mas (see Table~2). This implies a lower limit of
$\langle\theta\rangle = 1.0$~mas, if measurement with 1\% accuracy
is required. Angular diameters of 13 Cepheids are above this
limit, four of which have not been yet observed (SV~Vul, U~Car,
RS~Pup and overtone pulsator FF~Aql). Measuring
$\langle\theta\rangle$ with 2\% precision is achievable for
additional 39 stars.

Most interesting for interferometric observations are those
Cepheids, whose angular diameter {\it variations} can be detected.
Such a feat has been possible for stars with $\Delta\theta >
0.15$\th mas ({\it cf.} Table~2). 13 Cepheids satisfy this
condition. These variables cover uniformly the period range of
$\log P = 0.73 - 1.65$ (see Fig.\th\ref{f3}) and as such, are very
well suited for calibration of Cepheid $P-L$ and $P-R$ relations.
So far, only for five of them angular diameter variations have been
measured. The remaining eight Cepheids are SV~Vul, U~Car, RS~Pup,
T~Mon, X~Cyg, $\delta$~Cep, RZ~Vel, and TT~Aql. Their pulsations
can be resolved with accuracy already demonstrated by existing
interferometers. We encourage observers to concentrate their
efforts on these objects.

\Section{Conclusions}

Optical/near infrared long-baseline interferometry offers new ways
of studying Cepheid pulsations. With the goal of aiding such
studies, we calculated expected mean angular diameters and
amplitudes of angular diameter variations for all monoperiodic
Population~I Cepheids brighter than $\langle V\rangle = 8.0$\th
mag. Distances to the stars and their mean linear radii were
estimated with $P-L$ ($V$-band) and $P-R$ relations, respectively.
The amplitudes of radius variations were calculated precisely, by
integrating the observed radial velocity curves. Resulting mean
angular diameters and angular diameter amplitudes are listed in
Table~1. This catalog is intended to serve as a planning tool for
future interferometric observations of Cepheids.

Of particular interest are Cepheids, in which angular diameter
changes associated with pulsations can be detected. For such
stars, distances and linear radii can be determined by purely
geometrical Baade-Wesselink method ({\it e.g.} Lane et~al. 2000;
Kervella et~al. 2004a). This approach combines measured angular
diameter variations with linear displacement of the photosphere,
inferred by integrating the observed radial velocity curve.

Determination of {\it mean} angular diameter, which is possible
for many more Cepheids, is also of great interest. Such
measurements combined with accurately calibrated period--radius
relation can yield distances to low amplitude or far away
Cepheids, whose angular diameter changes are too small to be
detected. This will vastly enlarge the sample of objects available
for calibrating Cepheid $P-L$ relation.

Measuring angular diameters for many Cepheids covering widest
possible range of effective temperatures is also invaluable for
precise calibration of surface brightness-colour relations
(Nordgren et~al. 2002; Kervella et~al. 2004c). These relations are
a cornerstone of the near-IR Barnes-Evans method, which has a
potential of measuring accurate distances even to Cepheids in the
Magellanic Clouds ({\it e.g.} Storm et~al. 2004; Gieren et~al.
2005).

We identified 13 Cepheids with angular diameter amplitudes large
enough to be measured with precision already provided by
VINCI/VLTI and PTI interferometers. For seven of them, no
interferometric observations exist and for the eighth one
($\delta$~Cep) only mean angular diameter was published. Five of
these stars are easily accesible to VLTI facility.

With VLTI/AMBER interferometer (baseline 202m) coming into service
in 2005 and CHARA array already in operation (baseline of 330m),
the number of Cepheids accesible to interferometric study will
sharply increase. Currently, the magnitude limit for both
instruments is $K\sim 6$mag (Sturmann et.~al. 2003; AMBER
Commisioning~2 preliminary report,
http://www-laog.obs.ujf-grenoble.fr/amber/). This places almost
all Cepheids of our sample within reach. Both AMBER and CHARA can
work not only in $K$-band ($\lambda = 2.18\mu$m), but also in
$H$-band ($\lambda = 1.65\mu$m). With shorter wavelength and
longer baseline the expected angular resolution will increase 2-3
times, as compared to VINCI/VLTI performance. Consequently, the
number of Cepheids, for which angular diameter variations (hence
geometrical distances) can be measured will more than double,
reaching $\sim 30$. The {\it mean} angular diameters could be
determined to 1\% precision in more than 50 Cepheids and to 2\%
precision in all Cepheids of our sample. With planned extension of
CHARA and AMBER capabilities to $J$-band and eventually to
$V$-band, further increase in resolution is expected in the
future. The list of Cepheids with interferometrically detectable
pulsations will continue growing longer, creating excellent
prospect for very accurate calibration of Cepheid $P-L$ and $P-R$
relations.

\bigskip

\noindent {\bf Acknowledgements} This work has been supported in
part by KBN grant 5 P03D 030 20.

\end{document}